\pdfoutput=1
\documentclass[a4paper,11pt]{article}
\usepackage{amsfonts}
\usepackage{amsmath}
\usepackage{natbib}
\usepackage{comment}
\usepackage{graphicx}
\usepackage{subcaption}
\usepackage{multirow}
\usepackage{float}
\usepackage{hyperref}
\usepackage[margin=1.5in]{geometry}
\providecommand{\keywords}[1]{\textbf{Keywords:} #1}
\renewcommand\footnotemark{}

\title{A sequential reduction method for inference in generalized linear mixed models}
\author{Helen Ogden}
\date{University of Warwick, Coventry, UK  \\ \url{warwick.ac.uk/heogden}}

\begin{document}
\maketitle
\begin{abstract}
The likelihood for the parameters of a generalized linear mixed model involves an integral
 which may be of very high dimension. Because of this intractability,
many approximations to the likelihood have been proposed, but all can fail when the model is sparse, in that there is only a small amount of information available on each random effect.
  The sequential reduction method described in this paper exploits the dependence structure of the posterior distribution of the random effects to reduce substantially the cost of finding an accurate approximation to the
  likelihood in models with sparse structure.
\end{abstract}
\keywords{Graphical model, Intractable likelihood, Laplace approximation, Pairwise comparisons, Sparse grid interpolation}

\section{Introduction}
Generalized linear mixed models are a natural and widely used class of models, but one in which the likelihood often
involves an integral of very high dimension. Because of this intractability, many alternative methods
have been developed for inference in these models.

One class of approaches involves replacing the likelihood with some
approximation, for example using Laplace's method or importance sampling. However, these approximations
can fail in cases where the structure of the model is sparse, in that only a small amount of information
is available on each random effect, especially when the data are binary.

If there are $n$ random effects in total, the likelihood may always be written as an $n$-dimensional integral over
these random effects. If there are a large number of random effects, then it will be computationally infeasible
 to obtain an accurate
approximation to this $n$-dimensional
integral by direct numerical integration. However, it is not always necessary to compute this $n$-dimensional integral
to find the likelihood.
In a two-level random intercept model, independence between clusters may be exploited to write the likelihood as a product of
$n$ one-dimensional integrals, so it is relatively easy to obtain a good approximation to the likelihood, even for large $n$.
 In more complicated situations it is often not immediately obvious whether any such simplification exists.

The `sequential reduction' method developed in this paper exploits the structure of the integrand
to simplify computation of the likelihood, and as a result allows a fast and accurate approximation to the likelihood
to be found in many cases where existing approximation methods fail.
Examples are given to demonstrate the new method, including pairwise competition models and a model
with nested structure.

\section{The generalized linear mixed model}
\subsection{The model}
A generalized linear model \citep{Nelder_Wedderburn_1972} allows the distribution of a response $\mathbf{Y} = (Y_1, \ldots, Y_m)$ to depend on observed covariates
through a linear predictor $\mathbf{\eta}$, where
$\mathbf{\eta} = X \mathbf{\beta},$
for some known design matrix $X$. Conditional on knowledge of the linear predictor,
 the components of $\mathbf{Y}$ are independent. The distribution of $\mathbf{Y}$ is assumed to have exponential
family form, with mean
$\mu = \mathbb{E}(\mathbf{Y}|\eta) = g^{-1}(\eta),$
for some known link function $g(.)$.

An assumption implicit in the generalized linear model is that the distribution of the response is entirely
determined by the values of the observed covariates. In practice, this assumption is rarely believed: in fact,
there may be other information not encoded in the observed covariates which may affect the response.
A generalized linear mixed model allows for this extra heterogeneity by modeling the linear predictor as
$\mathbf{\eta} = X \mathbf{\beta} + Z(\psi) \mathbf{u},$
where $\mathbf{u}= (u_1,\ldots, u_n)$, and the $u_i$ are independent samples from some known distribution.
This paper concentrates on the case $u_i \sim N(0,1)$, which allows $Z(\psi) \mathbf{u}$
to have any multivariate normal distribution with mean zero.

The non-zero elements of the columns of $Z(\psi)$ give us the observations which involve each
random effect. We will say the generalized linear mixed model has `sparse structure' if most of
these columns have few non-zero elements, so that most random effects are only involved in
a few observations. These sparse models are particularly problematic for inference, especially
 when the data are binary, because the amount of information available on each random effect is small.

\subsection{Example: pairwise competition models}
\label{sec:pair_competition}
Consider a tournament among $n$ players, consisting of contests between pairs of players.
For each contest, we observe a binary outcome: either $i$ beats $j$ or $j$ beats $i$.
We suppose that each player $i$ has some ability $\lambda_i$, and that
conditional on all the abilities, the outcomes of the contests are independent, with distribution
depending on the difference in abilities of the players $i$ and $j$, so that
$\text{Pr}(\text{$i$ beats $j$}|\lambda) = g^{-1}(\lambda_i-\lambda_j)$
for some link function $g(.)$.
If $g(x)=\text{logit} (x)$, then this
describes a Bradley-Terry model \citep{Bradley_Terry_1952}. If $g(x) = \Phi^{-1}(x)$  (the probit link), then it
describes a Thurstone-Mosteller model \citep{Thurstone_1927,Mosteller_1951}.

If covariate information $\mathbf{x}_i$ is available for each player, then interest may lie in the effect of the observed covariates on ability, rather than the individual
abilities $\lambda_i$ themselves.
We allow the ability of player $i$ to depend on the covariates
$\mathbf{x}_i$ through $\lambda_i = \beta^T \mathbf{x}_i + \sigma u_i$,
where $u_i$ are independent $N(0,1)$ samples. This gives
 a generalized linear mixed model, depending on a linear predictor
$\eta$ with components $\eta_r = \lambda_{p_1(r)}-\lambda_{p_2(r)}$,
 where $p_1(r)$ and $p_2(r)$ are the first and second player involved in match $r$.
The model will have sparse structure if each player competes in only a small number of matches,
which is a common scenario in practice.


\subsection{The likelihood}
Let $f(.|\eta_i)$ be the density of $Y_i$, conditional on knowledge of the value of $\eta_i$,
and write $\theta=(\beta,\psi)$ for the full set of model parameters.
Conditional on $\eta$, the components of $\mathbf{Y}$ are independent, so that
\begin{equation}
L(\theta) = \int_{\mathbb{R}^n} \prod_{i=1}^m f\left(y_i|\eta_i= X_i^T \mathbf{\beta} + Z_i(\psi)^T \mathbf{u} \right) \prod_{j=1}^n \phi(u_j) du_j,
\label{eqn:glmm_likelihood}
\end{equation}
where $X_i$ is the $i$th row of $X$, and $Z_i(\psi)$ is the $i$th row
of $Z(\psi)$.
Unless $n$ is very small, it will not be possible to approximate the likelihood well by direct
computation of this $n$-dimensional integral.

\subsection{Existing approximations to the likelihood}
\label{sec:laplace}

\cite{Pinheiro_Bates_1995} suggest using a Laplace approximation to the integral \eqref{eqn:glmm_likelihood}.
Write
\begin{equation*}
g(u_1, \ldots, u_n| \mathbf{y}, \theta) = \prod_{i=1}^m f\left(y_i|\eta_i= X_i^T \mathbf{\beta} + Z_i(\psi)^T \mathbf{u} \right) \prod_{j=1}^n \phi(u_j)
\end{equation*}
for the integrand of the likelihood. This may be thought of as a non-normalized version of the posterior
density for $\mathbf{u}$, given $\mathbf{y}$ and $\theta$.
For each fixed $\theta$, the Laplace approximation relies on a normal approximation to this posterior density. To find this normal approximation,
let $\mu_\theta$ maximize $\log g(\mathbf{u}| \mathbf{y}, \theta)$ over
$\mathbf{u}$, and write $\Sigma_\theta = -H_\theta^{-1}$, where $H_\theta$ is the Hessian
resulting from this optimization.  The normal approximation to $g(.|\mathbf{y}, \theta)$ will be proportional to a
$N_n(\mu_\theta,\Sigma_\theta)$ density. Writing $g^\text{na}(.|\mathbf{y}, \theta)$ for the normal approximation to $g(.|\mathbf{y}, \theta)$,
\[g^\text{na}(\mathbf{u}|\mathbf{y}, \theta) = \frac{g(\mu_\theta|\mathbf{y},\theta)}{\phi_n(\mu_\theta; \mu_\theta,\Sigma_\theta)} \phi_n(\mathbf{u};\mu_\theta,\Sigma_\theta),\]
where we write $\phi_n(.;\mu,\Sigma)$ for the $N_n(\mu,\Sigma)$ density.
When we integrate over $\mathbf{u}$, only the normalizing constant remains, so that
\[L^\text{Laplace}(\theta) =  \frac{g(\mu_\theta|\mathbf{y}, \theta)}{\phi_n(\mu_\theta;\mu_\theta,\Sigma_\theta)}
= (2 \pi)^{-\frac{n}{2}} (\det \Sigma_\theta)^{-\frac{1}{2}} g(\mu_\theta|\mathbf{y}, \theta). \]

In the case of a linear mixed model, the approximating normal density is precise, and there is no
error in the Laplace approximation to the likelihood. In other cases, and particularly when the response is discrete and
may only take a few values, the error in the Laplace approximation may be large.
 In the case that
$n$ is fixed, and $m \rightarrow \infty$, the relative error
in the Laplace approximation may be shown to tend to zero.
However, in the type of model we consider here, $n$ is not fixed, but grows with $m$.
The validity of the Laplace approximation depends upon the rate of this growth.
\cite{Shun_McCullagh_1995} study this problem, and conclude that
the Laplace approximation should be reliable provided that
$n=o(m^{1/3})$.
However, the Laplace approximation to the difference in the log-likelihood at two nearby points
tends to be much more accurate than the approximation to the log-likelihood itself.
The effect that ratios of Laplace approximations to similar functions tend to be
more accurate than each Laplace approximation individually has been noted before, for example by \cite{tierney_kadane_1986} in the context of
computing posterior moments.
Nonetheless, in models with very sparse structure (where we might have $n=O(m)$), even the shape of
the Laplace approximation to the log-likelihood surface may be inaccurate, so another method is required.

In cases where the Laplace approximation fails, \cite{Pinheiro_Bates_1995} suggest
constructing an importance sampling approximation to the likelihood, based on
samples from the normal distribution $N_n(\mu_\theta,\Sigma_\theta)$. Writing
\[w(\mathbf{u};\theta) = \frac{g(\mathbf{u}|\theta)}{\phi_n(\mathbf{u};\mu_\theta,\Sigma_\theta)},\]
the likelihood may be approximated by
$L^{IS}(\theta) = \sum_{i=1}^N w(\mathbf{u}^{(i)};\theta)/N,$
where $\mathbf{u}^{(i)} \sim N(\mu_\theta,\Sigma_\theta)$.

Unfortunately, there is no guarantee that the variance of the importance weights $w(\mathbf{u}^{(i)};\theta)$ will be finite.
 In such a situation,
 the importance sampling approximation
will still converge to the true likelihood as $N \rightarrow \infty$, but the convergence may be slow and
erratic, and estimates of the variance of the approximation may be unreliable.

\subsection{Bayesian inference}

From a Bayesian perspective, Markov chain Monte Carlo methods could be used
to sample from the posterior distribution. However, such methods
are computationally intensive, and it can be difficult to detect whether the Markov chain
has converged to the correct distribution.
\cite{Rue_etal_2009} suggest the Integrated Nested Laplace Approximation (INLA)
 to approximate
the marginal posterior distribution of each parameter.
INLA is computationally efficient, but \cite{Fong_etal_2010} note that the approximation may
perform poorly in models for binary data. In situations where the Laplace approximation
to the likelihood fails, INLA may be also unreliable.

We do not consider these methods further, and instead focus on
those methods which provide a
direct approximation to the (marginal) likelihood \eqref{eqn:glmm_likelihood}.

\section{The sequential reduction method}
\label{sec:SR}
\subsection{Conditional independence structure}
Before observing the data $\mathbf{y}$, the random effects $\mathbf{u}$ are independent.
The information provided by $\mathbf{y}$ about the value of combinations of
those random effects induces dependence between them. If there is no observation involving both $u_i$ and $u_j$,
$u_i$ and $u_j$ will be conditionally independent
in the posterior distribution, given the values of all the other random effects.

It is possible to represent this conditional independence structure graphically.
Consider a graph $\mathcal{G}$ constructed to have:
\begin{enumerate}
\item{A vertex for each random effect}
\item{An edge between two vertices if there is at least one observation involving both of the corresponding random effects.}
\end{enumerate}
By construction of $\mathcal{G}$, there is an edge between $i$ and $j$ in $\mathcal{G}$ only if $\mathbf{y}$ contains an observation involving both $u_i$ and $u_j$.
So if there is no edge between $i$ and $j$ in $\mathcal{G}$,
$u_i$ and $u_j$ are conditionally independent
in the posterior distribution, given the values of all the other random effects, so the
posterior distribution of the random effects has
the pairwise Markov property with respect to $\mathcal{G}$.
We call $\mathcal{G}$ the posterior dependence graph for
$\mathbf{u}$ given $\mathbf{y}$.

In a pairwise competition model, the posterior dependence graph simply consists of a vertex for each player, with
an edge between two vertices if those players compete in at least one contest. For models in which each observation relies
on more than two random effects, an observation will not be represented by a single edge in the graph.

The problem of computing the likelihood has now been transformed to that of finding a normalizing
constant of a density associated with an undirected graphical model.
In order to see how the conditional dependence structure can be used to enable a simplification
of the likelihood, we first need a few definitions.
A complete graph is one in which there is an edge from each vertex to every other vertex.
A clique of a graph $\mathcal{G}$ is a complete subgraph of $\mathcal{G}$, and a clique
is said to be maximal if it is not itself contained within a larger clique.
For any graph $\mathcal{G}$, the set of all maximal
cliques of $\mathcal{G}$ is unique, and we write $M(\mathcal{G})$ for
this set.

The Hammersley-Clifford theorem \citep{Besag_1974} implies that
$g(.|\mathbf{y},\theta)$ factorizes over the maximal cliques of $\mathcal{G}$, so that we may write
\[g(\mathbf{u}|\mathbf{y},\theta) = \prod_{C \in M(\mathcal{G})} g_C(\mathbf{u}_C)\]
for some functions $g_C(.)$.
A condition needed to obtain this result using the Hammersley-Clifford theorem is that
$g(\mathbf{u}|\mathbf{y},\theta)>0$ for all $\mathbf{u}$. This will hold in this case because $\phi(u_i)>0$
for all $u_i$. In fact, we may show that such a factorization exists directly. One particular such
factorization is constructed in Section \ref{sec:factorization_construction}, and would be valid even if
we assumed a random effects density $f_u(.)$ such that $f_u(u_i)=0$ for some $u_i$.

\subsection{Exploiting the clique factorization}
\cite{Jordan_2004} reviews some methods to find the marginals of a
density factorized over the maximal cliques of a graph.
While these methods are well known,
their use is typically limited to certain special classes of distribution, such as discrete or Gaussian distributions.
We will
use the same ideas, combined with a method for approximate storage of functions, to approximate the marginals
of the distribution with density proportional to $g(.|\mathbf{y},\theta)$, and so
approximate the likelihood
$L(\theta) = \int_{\mathbb{R}^n} g(\mathbf{u}|\mathbf{y},\theta) d\mathbf{u}.$

We take an iterative approach to the problem, first integrating out $u_1$
to find the non-normalized marginal posterior density of $\{u_2, \ldots, u_n\}$.
We start with a factorization of $g(.|\mathbf{y},\theta)$ over the maximal cliques of the posterior dependence graph of
$\{u_1, \ldots, u_n\}$,
and the idea will be to write the marginal posterior density of $\{u_2, \ldots, u_n\}$ as a product
over the maximal cliques of a new marginal posterior dependence graph. Once this is done, the process
may be repeated $n$ times to find the likelihood.
We will write $\mathcal{G}_i$ for the posterior dependence graph of $\{u_i,\ldots,u_n\}$, so we start
with posterior dependence graph $\mathcal{G}_1=\mathcal{G}$. Write $M_i = M(\mathcal{G}_i)$ for
the maximal cliques of $\mathcal{G}_i$.

Factorizing $g(.|\mathbf{y},\theta)$ over the maximal cliques of $\mathcal{G}_1$ gives
\[g(\mathbf{u}|\mathbf{y},\theta) = \prod_{C \in M_1} g_C^{1}(\mathbf{u}_C),\]
for some functions $\{g_C^{1}(.): C \in M_1\}$.
To integrate over $u_1$, it is only necessary to integrate over
 maximal cliques containing vertex $1$, leaving the functions on other cliques unchanged.
 Let $N_1$ be the set of neighbors of vertex $1$ in $\mathcal{G}$ (including vertex $1$ itself).
Then
 \begin{align*}
 \int g(\mathbf{u}|\mathbf{y},\theta) du_1 &= \int \prod_{C \in M_1: C \subseteq N_1}  g^{1}_C(\mathbf{u}_C) du_1 \prod_{\tilde C \in M_1: \tilde C \not\subseteq N_1} g^{1}_{\tilde C}(\mathbf{u}_{\tilde C}) \\
 &= \int g^{1}_{N_1}(u_1, \mathbf{u}_{N_1 \setminus 1}) du_1  \prod_{\tilde C \in M_1: \tilde C \not\subseteq N_1} g^{1}_{\tilde C}(\mathbf{u}_{\tilde C}).
 \end{align*}
 Thus $g^{1}_{N_1}(.)$ is obtained by multiplication of all the functions on
 cliques which are subsets of $N_1$. This is then integrated over $u_1$, to give
 \[g^{2}_{N_1 \setminus 1} (\mathbf{u}_{N_1 \setminus 1}) = \int g^{1}_{N_1}(u_1, \mathbf{u}_{N_1 \setminus 1}) du_1. \]
The functions on all cliques $\tilde C$ which are
 not subsets of $N_1$ remain unchanged, with
 $g^{2}_{\tilde C} (\mathbf{u}_{\tilde C})=g^{1}_{\tilde C} (\mathbf{u}_{\tilde C})$.

This defines a new factorization of $g(u_2, \ldots u_n|\mathbf{y},\theta)$ over the maximal cliques $M_2$
of the posterior dependence graph for $\{u_2, \ldots, u_n\}$, where $M_2$ contains $N_1 \setminus 1$,
and all the remaining cliques in $M_1$ which are not subsets of $N_1$.
The same process may then be followed to remove each $u_i$ in turn.
\subsection{The sequential reduction method}
\label{sec:SR_alg_general}

We now give the general form of a
 sequential reduction method for approximating the likelihood. We highlight the places
 where choices must be made to use this method in practice.
The following sections then discuss each of these choices in detail.

\begin{enumerate}
\item{The $u_i$ may be integrated out in any order. Section \ref{sec:cost} discusses
how to choose a good order, with the aim of minimizing the cost
of approximating the likelihood. Reorder the random effects so that we integrate out $u_1, \ldots, u_n$
in that order.}
\item{Factorize $g(\mathbf{u}|\mathbf{y},\theta)$ over the maximal cliques $M_1$ of the posterior
dependence graph, as
$g(\mathbf{u}|\mathbf{y},\theta) = \prod_{C \in M_1} g^{1}_C(\mathbf{u}_C).$
This factorization is not unique, so we must choose one particular factorization $\{ g^{1}_C(.): C \in M_1\}$.
Section \ref{sec:factorization_construction} gives the factorization we use in practice.
}
\item{
\label{step:integrate}
Once $u_1, \ldots u_{i-1}$ have been integrated out (using some approximate method), we have the factorization
$\tilde g(u_i,\ldots,u_n | \mathbf{y},\theta) = \prod_{C \in M_i} g^{i}_C(\mathbf{u}_C),$
of the (approximated) non-normalized posterior for $u_i, \ldots, u_n$.
Write
\[g_{N_i}(\mathbf{u}_{N_i}) = \prod_{C \in M_i: C \subset N_i } g^{i}_C(\mathbf{u}_C).\]
We then integrate over $u_i$ (using a quadrature rule), and store an approximate representation $\tilde g_{N_i \setminus i}(.)$
of the resulting function
$g_{N_i \setminus i}(.)$.
In Section \ref{sec:storage} we discuss the construction of this approximate
representation.
}
\item{
\label{step:redefine_pdg}
Write
 \[\tilde g(u_{i+1},\ldots,u_n | \mathbf{y},\theta) = \tilde g_{N_i \setminus i}(\mathbf{u}_{N_i \setminus i}) \prod_{C \in M_i: C \not \subset N_i } g^{i}_C(\mathbf{u}_C), \]
defining a factorization of the (approximated) non-normalized posterior density of $\{u_{i+1},\ldots,u_n\}$
over the maximal cliques $M_{i+1}$
of the new posterior dependence graph $\mathcal{G}_{i+1}$.
}
\item{Repeat steps \eqref{step:integrate} and \eqref{step:redefine_pdg} for $i=1,\ldots,n-1$, then
integrate $\tilde g(u_n|\mathbf{y},\theta)$ over $u_n$ to give the approximation to the likelihood.}
\end{enumerate}

\subsection{A specific clique factorization}
\label{sec:factorization_construction}

The general method described in Section \ref{sec:SR_alg_general} is valid for an arbitrary factorization of $g(\mathbf{u}|\mathbf{y},\theta)$
over the maximal cliques $M_1$ of the posterior dependence graph.
To use the method in practice, we must first define the factorization used.

Given an ordering of the vertices, order the cliques in $M_1$ lexicographically according to the
set of vertices contained within them. The observation vector $\mathbf{y}$ is partitioned over the cliques in $M_1$
by including in $\mathbf{y}_C$ all the observations only involving items in the clique $C$, which
have not already been included in $\mathbf{y}_B$ for some earlier clique in the ordering, $B$.
Write $a(C)$ for the set of vertices appearing for the first time in clique $C$. Let
\[g^{1}_C(\mathbf{u}_C) = f(\mathbf{y}_C|\mathbf{u}_C) \prod_{j \in a(C)}\phi(u_j).\]
Then
$g(\mathbf{u}|\mathbf{y}) = \prod_{C \in M_1} g^{1}_C(\mathbf{u}_C),$
so $g^{1}_C(.)$ does define a factorization of $g(.|y)$.

\subsection{Approximate function representation}
\label{sec:storage}
\subsubsection{A modified function for storage}
A key choice in the sequential reduction algorithm is the method used to `store' the function
$g_{N_i \setminus i}(.)$. The storage consists of a set of points $S_i$ at which to
evaluate $g_{N_i \setminus i}(.)$, and a method of interpolation between those points,
which will be used later in the algorithm if we need to evaluate
$g_{N_i \setminus i}(\mathbf{u}_{N_i \setminus i})$ for some $\mathbf{u}_{N_i \setminus i} \not \in S_i$.

We would like to minimize the size of the absolute error in the
interpolation for those points $\mathbf{u}_{N_i \setminus i}$
at which we will later interpolate.
The quality of the interpolation may be far more important at some points
$\mathbf{u}_{N_i \setminus i}$ than at others.
We will transform to a new function
$r_{N_i \setminus i}(\mathbf{u}_{N_i \setminus i}) = g_{N_i \setminus i}(\mathbf{u}_{N_i \setminus i})h_{N_i \setminus i}(\mathbf{u}_{N_i \setminus i}),$
where we choose $h_{N_i \setminus i}(.)$
so that the size of the absolute interpolation error for
$r_{N_i \setminus i}(.)$ is of roughly equal concern across the whole space.
Given an interpolation method for $r_{N_i \setminus i}(.)$, we obtain interpolated values
for  $g_{N_i \setminus i}(.)$ through
$g_{N_i \setminus i}^\text{interp}(\mathbf{u}_{N_i \setminus i}) = r^\text{interp}_{N_i \setminus i}(\mathbf{u}_{N_i \setminus i})/h_{N_i \setminus i}(\mathbf{u}_{N_i \setminus i}),$ so we must ensure that
$h_{N_i \setminus i}(.)$ is easy to compute.

Recall that we may think of the original integrand $g(.|\mathbf{y},\theta)$ as being
the non-normalized posterior density for $\mathbf{u}|\mathbf{y},\theta$. The region
where where we will interpolate a large number of points
 corresponds to the region where the marginal posterior density of
$\mathbf{u}_{N_i \setminus i}|\mathbf{y},\theta$ is large.
Ideally, we would choose $h_{N_i \setminus i}(.)$ to make
 $r_{N_i \setminus i}(.)$ proportional to the density of
  $\mathbf{u}_{N_i \setminus i} | \mathbf{y},\theta$, but this density is difficult to compute.

To solve this problem, we make use of the normal approximation to $g(.|\mathbf{y},\theta)$
used to construct the Laplace approximation to the likelihood, which approximations
the posterior distribution $\mathbf{u}|\mathbf{y},\theta$ as $N_n(\mu,\Sigma)$. The marginal posterior
distribution of $\mathbf{u}_{N_i \setminus i}|\mathbf{y},\theta$ may therefore be
approximated as $N_d(\mu_{N_i\setminus i},\Sigma_{N_i\setminus i}),$ where $d=|N_i \setminus i|$.
We choose $h_{N_i\setminus i}(.)$ to ensure that the normal approximation to $r_{N_i\setminus i}(.)$ (computed as
described in Section \ref{sec:laplace})
is $N_d(\mu_{N_i\setminus i},\Sigma_{N_i\setminus i})$.
That is, we choose $\log h_{N_i\setminus i}(.)$ to be a quadratic function, with coefficients chosen so that
$\nabla \log h_{N_i\setminus i}(\mu_{N_i\setminus i}) = -\nabla \log g_{N_i\setminus i}(\mu_{N_i\setminus i})$
and
$\nabla^T \nabla \log h_{N_i\setminus i}(\mu_{N_i\setminus i})
= -\Sigma_{N_i\setminus i}^{-1} -\nabla^T \nabla \log g_{N_i\setminus i}(\mu_{N_i\setminus i}).$

\subsubsection{Storing a function with a normal approximation}
Suppose that $f(.)$ is a non-negative function on $\mathbb{R}^d$, for which we want to store an approximate representation,
and that we may approximate $f(.)$ with $f^\text{na}(\mathbf{x}) \propto \phi_d(\mathbf{x},\mu,\Sigma)$, for some $\mu$ and $\Sigma$.
In our case, the function $f(.)$ which we store is $r_{N_i\setminus i}(.)$, of dimension $d=|N_i \setminus i|$, and
with normal approximation $N_d(\mu_{N_i \setminus i}, \Sigma_{N_i \setminus i})$.

We transform to a new basis. Let $\mathbf{z} = A^{-1}(\mathbf{x}-\mu)$, where $A$ is chosen so that $A A^T=\Sigma$.
More specifically, we choose $A=P D$, where $P$ is a matrix whose columns are the normalized eigenvectors of
$\Sigma$ and $D$ is a diagonal matrix with diagonal entries the square roots of the eigenvalues of $\Sigma$.
 Write $f_z(\mathbf{z}) = f(A \mathbf{z} + \mu)$, and let $c(\mathbf{z}) = \log f_z(\mathbf{z}) - \log \phi_d(\mathbf{z},0,I)$, so that $c(.)$ will be constant
if the normal approximation is precise. We store $c(.)$ by evaluating at some fixed points for $\mathbf{z}$,
and specifying the method
of interpolation between them. The choice of these points and the interpolation method
is discussed in the next section. Given the interpolation method for $c(.)$, we may define
$f^\text{interp}(\mathbf{x}) = \exp\{ c^\text{interp}(A^{-1}(\mathbf{x}-\mu))\} \, \phi_d(A^{-1}(\mathbf{x}-\mu),0,I),$
to give an interpolation method for $f(.)$.

If $g(\mathbf{u}|\mathbf{y},\theta) \propto \phi_n(\mathbf{u},\mu,\Sigma)$, there will be no error in the
 Laplace approximation to the likelihood. In this situation, $c(.)$ will be constant, and the sequential reduction
approximation will also be exact.
In situations where the normal approximation is imprecise, $c(.)$ will no longer be constant, and
we may improve on the baseline (Laplace) approximation to the likelihood by increasing the number of points used for storage.

\subsubsection{Sparse grid interpolation}
\label{sec:interpolation}

In order to store an approximate representation of the standardized modifier function $c(.)$,
we will compute values of $c(.)$ at a fixed set of evaluation points, and specify a method of interpolation between these
points.
We now give a brief overview of the interpolation methods based on sparse grids of evaluation points.
Some of the notation we use is taken from
\cite{Barthelmann_etal_2000}, although there are some differences: notably that we assume
$c(.)$ to be a function on $\mathbb{R}^d$, rather than on the $d$-dimensional hypercube $[-1,1]^d$,
and we will use cubic splines, rather than
(global) polynomials for interpolation.

First we consider a method for interpolation for a one-dimensional function $c: \mathbb{R} \rightarrow \mathbb{R}$.
We evaluate $c(.)$ at $m_l$ points
$s_{1},\ldots,s_{m_l}$ and write
\[\mathcal{U}^l(c) =  \sum_{j=1}^{m_l} c(s_j) a_j^l,\]
where the $a_j^l$ are basis functions.   The approximate interpolated value of $c(.)$ at any point $x$ is then
given by $\mathcal{U}^l(c)(x)$.

Here $l$ denotes the level of approximation, and we suppose that the set of evaluation
points is nested so that at level $l$, we simply use the first $m_l$ points of a fixed set
of evaluation points $S=\{s_1, s_2, \ldots\}.$
We assume that $m_1=1$, so at the first level of approximation,
only one point is used, and $m_l=2^{l}-1$ for $l>1$, so there is an approximate doubling of the number
of points when the level of approximation is increased by one.

The full grid method of interpolation is to take
$m_{l_j}$ points in dimension $j$, and compute at each possible combination of those points. We write
\[ (\mathcal{U}^1\otimes \ldots \otimes \mathcal{U}^d)(c) = \sum_{j_1=1}^{m_{l_1}}\ldots \sum_{j_d=1}^{m_{l_d}} c(s_{j_1},...,s_{j_d}) \left( a^{l_1}_{j_1} \otimes \ldots \otimes a^{l_d}_{j_d}\right),\]
where
\[(a^{l_1}_{j_1} \otimes \ldots \otimes a^{l_d}_{j_d})(x_1,\ldots,x_d) = a^{l_1}_{j_1}(x_1)\times\ldots \times a^{l_d}_{j_d}(x_d).\]
Thus, in the full grid method, we must evaluate $c(.)$ at
$\prod_{j=1}^d {m_{l_j}} = O \left(\prod_{j=1}^d 2^{l_j} \right) = O\left(2^{\sum l_j}\right)$
  points. This will not be possible if $\sum_{j=1}^d l_j$ is too large.

In order to construct an approximate representation of $c(.)$ in reasonable time, we could limit the sum
$\sum_{j=1}^d l_j$ used in a full grid to be at most $d+k$, for some $k \geq 0$. If $k>0$,
there are many possibilities for `small full grids' indexed by the levels $\mathbf{l}=(l_1,\ldots, l_d)$
which satisfy this constraint.
A natural question is how
to combine the information given by each of these small full grids to give a good representation overall.

For a univariate function $c(.)$, let
\[\Delta^l(c) = \mathcal{U}^l(c)-\mathcal{U}^{l-1}(c)
= \sum_{j=1}^{m_{l-1}} c(s_j) \left[a_l^j - a_{l-1}^j \right] + \sum_{j=m_{l-1}+1}^{m_l} c(s_j) a_l^j,\]
for $l>1$, and $\Delta^1=U^1$. Then $\Delta^l$ gives the quantity we should add the approximate storage of $c(.)$ at
level $l-1$ to incorporate the new information given by the knots added at level $l$.

Returning to the multivariate case, the sparse grid interpolation of $c(.)$ at level $k$ is given by
\[ c^\text{interp}_k = \sum_{\mathbf{l}: |\mathbf{l}|\leq d+k} (\Delta^{l_1} \otimes \ldots \otimes \Delta^{l_d}) (c). \]
To store $c(.)$ on a sparse grid at level $k$, we must evaluate at $O\left(d^{k+1} \right)$
  points, which allows approximate storage for much larger dimension $d$
  than is possible using a full grid method.

\cite{Barthelmann_etal_2000} use global polynomial interpolation for
a function defined on a hypercube, with the Chebyshev knots. We prefer
to use cubic splines for interpolation, since the positioning of the knots is less critical.
Since we have already standardized the function we wish to store, we use the same knots in each direction, and
choose these standard knots $\mathbf{s}_l$ at level $l$ to be $m_l$ equally spaced quantiles
of a $N(0,\tau_k^2)$ distribution.
As $k$ increases, we choose larger $\tau_k$, so that the size of the region covered by the sparse grid increases with $k$.
However, the rate at which $\tau_k$ increases
should be sufficiently slow to ensure that the distance between the knots $\mathbf{s}_{k}$ decreases with $k$.
 Somewhat
arbitrarily, we choose $\tau_k=1+\frac{k}{2},$
which appears to work reasonably well in practice.

\subsubsection{Bounded interpolation}
To ensure that $g_{N_i}(.)$ remains integrable at each stage, we impose an upper bound $M$
on the interpolated value of $c(.)$. In practice, we choose $M$ to be the largest
value of $c(\mathbf{z})$ observed at any of the evaluation points.

\subsection{Computational complexity}
\label{sec:cost}

Using sparse grid storage at level $k$,
the cost of stage $i$ of the sequential reduction algorithm is at most $O(|N_i|^{2 k})$. The overall cost of approximating
the likelihood will be large if $\max_i |N_i|$
is large.

The random effects may be removed in any order, so it makes sense
to use an ordering that allows approximation of the likelihood at minimal cost. This problem may
 be reduced to a problem in graph theory: to find an ordering of the vertices
of a graph, such that when these nodes are removed in order, joining together all neighbors of the vertex
to be removed at each stage,
the largest clique obtained at any stage is as small as possible. This is known as the triangulation problem,
and the smallest possible value, over all possible orderings, of the largest clique obtained at some stage is known
as the treewidth of the graph.

Unfortunately, algorithms available to calculate the treewidth of a graph on $n$ vertices can take at worst $O(2^n)$
operations, so to find the exact treewidth may be too costly for $n$ at all large.
However, there are special structures of graph which have known treewidth, and algorithms exist to find
upper and lower bounds on the treewidth in reasonable time  \citep[see][]{treewidth_upper,treewidth_lower}.
We use a constructive algorithm for finding an upper bound on the treewidth, which outputs an elimination
ordering achieving that upper bound, to find a reasonably good (though not necessarily optimal) ordering.

\subsection{An R package for sequential reduction}
\label{sec:code}
The sequential reduction method is implemented in
R \citep{R} by the package glmmsr, which may be found at \url{warwick.ac.uk/heogden/code}.
The code for sparse grid interpolation is based on the efficient storage
schemes suggested by \cite{Murarasu_etal_2011}. Code
to reproduce the examples of Section \ref{sec:examples} is also provided.

\section{Examples}
\label{sec:examples}
We give some examples to compare the performance of the proposed
sequential reduction method with existing methods to approximate the likelihood.
The first two examples here are of pairwise competition models (a simple
tree tournament with simulated data, and a more complex, real-data example);
the third is a mixed logit model with two nested layers of random effects.

\begin{figure}
\centering
\begin{subfigure}[b]{0.3 \textwidth}
\includegraphics[width=\textwidth]{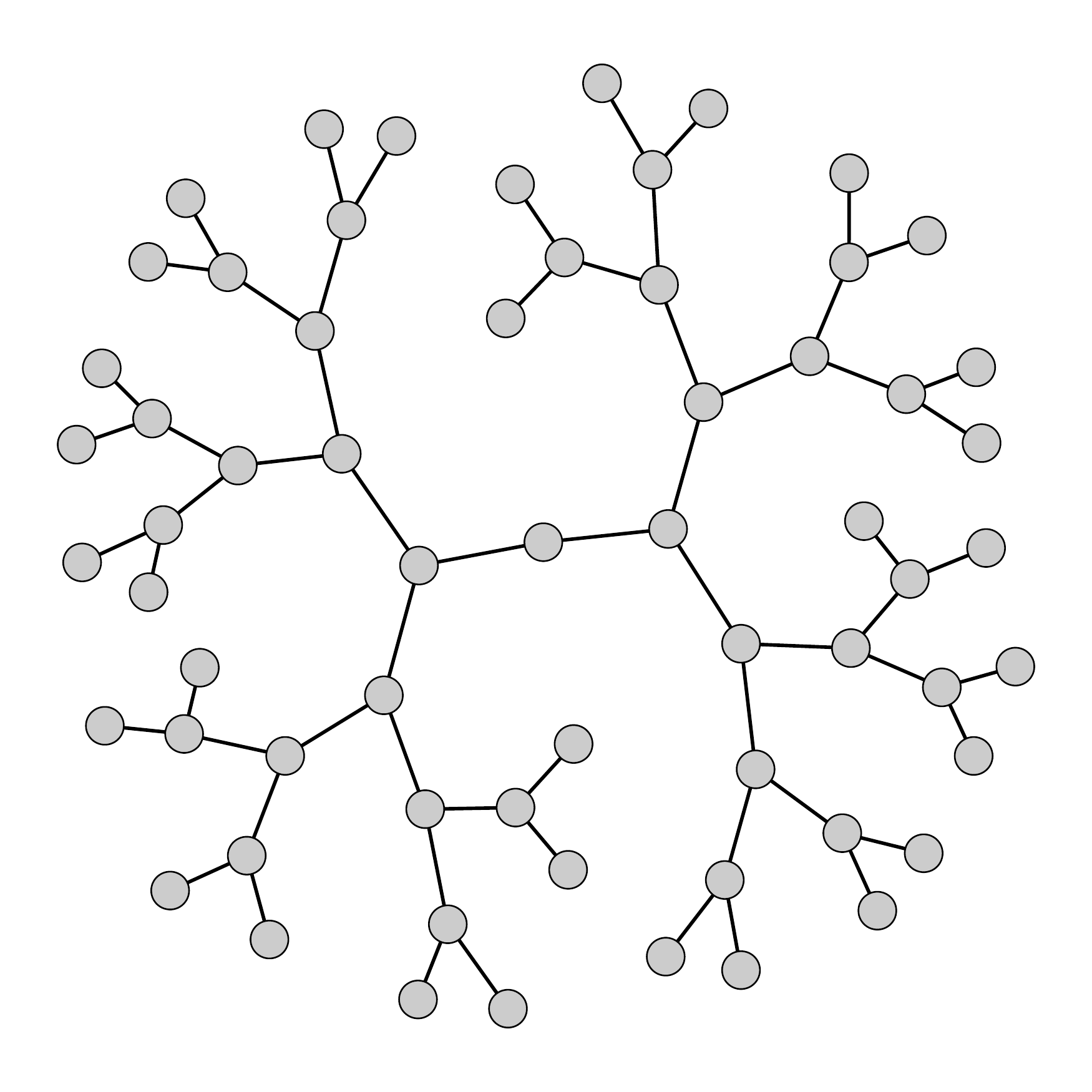}
 \caption{Tree tournament \label{fig:tree}}
 \end{subfigure}
 \begin{subfigure}[b]{0.3 \textwidth}
\includegraphics[width=\textwidth]{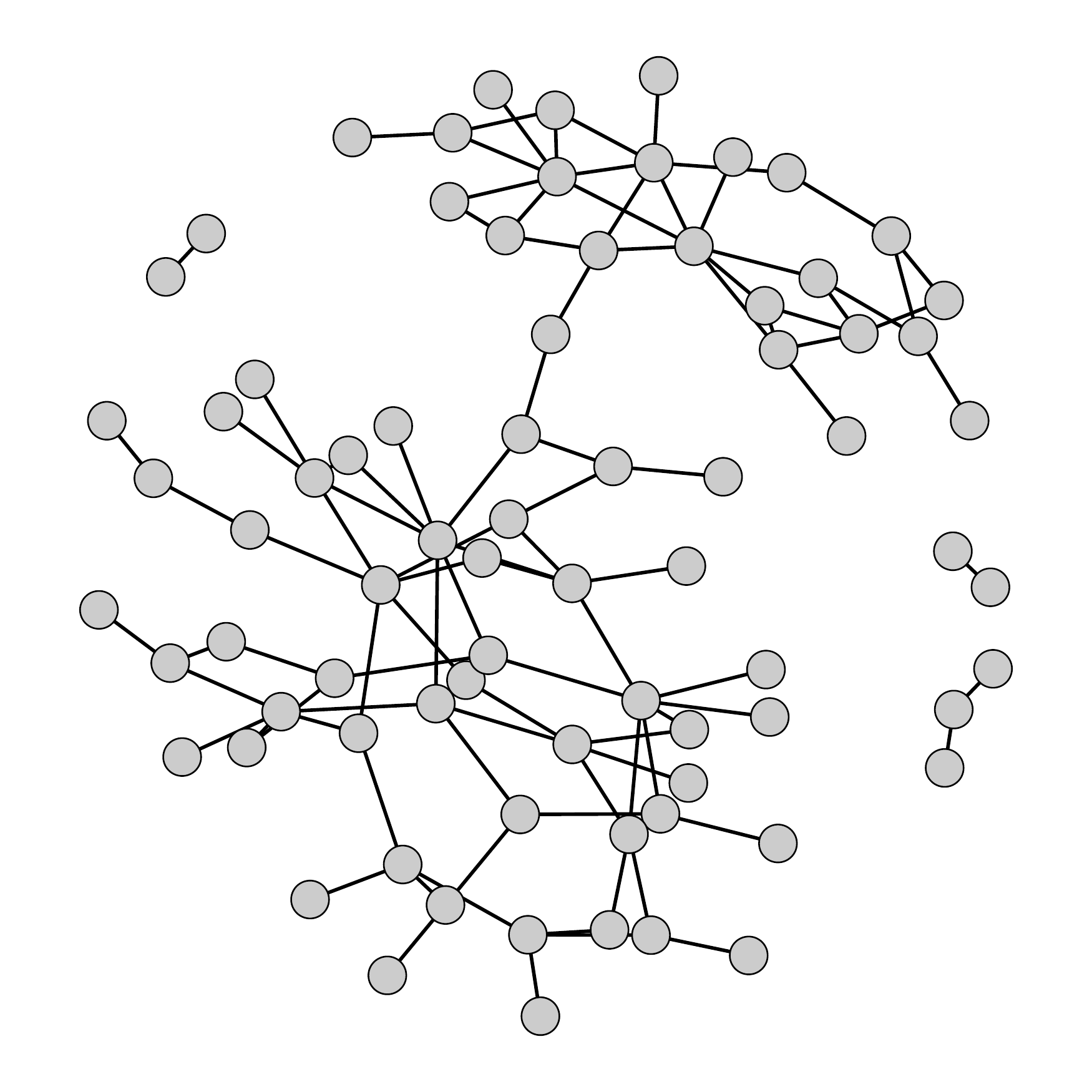}
 \caption{Lizards tournament\label{fig:lizards}}
\end{subfigure}
\begin{subfigure}[b]{0.3 \textwidth}
\setlength{\unitlength}{1cm}
\begin{picture}(5,5)
\put(1.4,3.2){\includegraphics[width=0.12\textwidth]{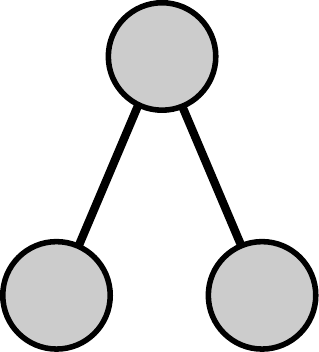}}
\put(1.4,2.2){\includegraphics[width=0.12\textwidth]{figures/three_level_each.pdf}}
\put(1.6,1.4){$\vdots$}
\put(1.4,0.4){\includegraphics[width=0.12\textwidth]{figures/three_level_each.pdf}}
\put(1.7,1.9){$\left. \begin{tabular}{l}  \\ \\ \\ \\ \\ \\ \end{tabular} \right\}$}
\put(2.6,1.9){$100$ times}
\end{picture}
 \caption{Three-level model\label{fig:three_level}}
\end{subfigure}
\caption{\label{fig:PDGs} The posterior dependence graphs for the examples}
\end{figure}

\subsection{Tree tournament}
\label{example:tree}

Consider observing a tree tournament, with structure as shown in Figure \ref{fig:tree}.
Suppose that there is a single observed covariate $x_i$ for each player, where $\lambda_i = \beta x_i + \sigma u_i$
and $u_i \sim N(0,1)$.
We consider one particular tournament with this tree structure, simulated from the model with
 $\beta=0.5$ and $\sigma=1.5$.
We suppose that we observe two matches between each pair of competing players.
The covariates $x_i$ are independent
draws from a standard normal distribution.

We fit the model using the Laplace approximation, and the sequential reduction approximations,
for $k=1$, $2$, $3$, $4$ and $5$.
 The posterior dependence graph of a tree tournament is a tree, which has treewidth $2$.
 Using the sequential reduction method with sparse grid storage at level $k$, the cost of approximating
the likelihood at each point will be $O(n 4^k)$. In reality, the computation time
does not quadruple each time $k$ is increased, since the computation is dominated by fixed operations
whose cost does not depend on $k$. To compute the approximation to the likelihood
at a single point took about $0.02$ seconds for the Laplace approximation, $0.22$ seconds for $k=1$,
$0.24$ seconds for $k=2$, $0.24$ seconds for $k=3$, $0.27$ seconds for $k=4$ and $0.30$ seconds for $k=5$.

Table \ref{table:tree_fit} gives the estimates of $\beta$ and $\sigma$ resulting from
each approximation to the likelihood. The estimates of $\beta$ are similar for all the approximations,
but the estimate of $\sigma$ found by maximizing the Laplace approximation to the likelihood
is smaller than the true maximum likelihood estimator.

\begin{table}[h]
\centering
\caption{\label{table:tree_fit} The parameter estimates and standard errors for the tree tournament}
\begin{tabular}{rrrrrrrr}
\hline
& & Laplace & k=1 & k=2 & k=3 & k=4 & k=5 \\
  \hline
\multirow{2}{*}{$\beta$} & estimate & 0.44 & 0.44 & 0.45 & 0.46 & 0.46 & 0.46 \\
 & s.e. & 0.26 & 0.27 & 0.27 & 0.27 & 0.27 & 0.27 \\
 \multirow{2}{*}{$\sigma$} & estimate & 1.13 & 1.26 & 1.29 & 1.30 & 1.30 & 1.30 \\
 &  s.e. & 0.31 & 0.36 & 0.37 & 0.38 & 0.38 & 0.38 \\
   \hline
\end{tabular}
\end{table}

\begin{figure}
\centering
\includegraphics[width=\textwidth]{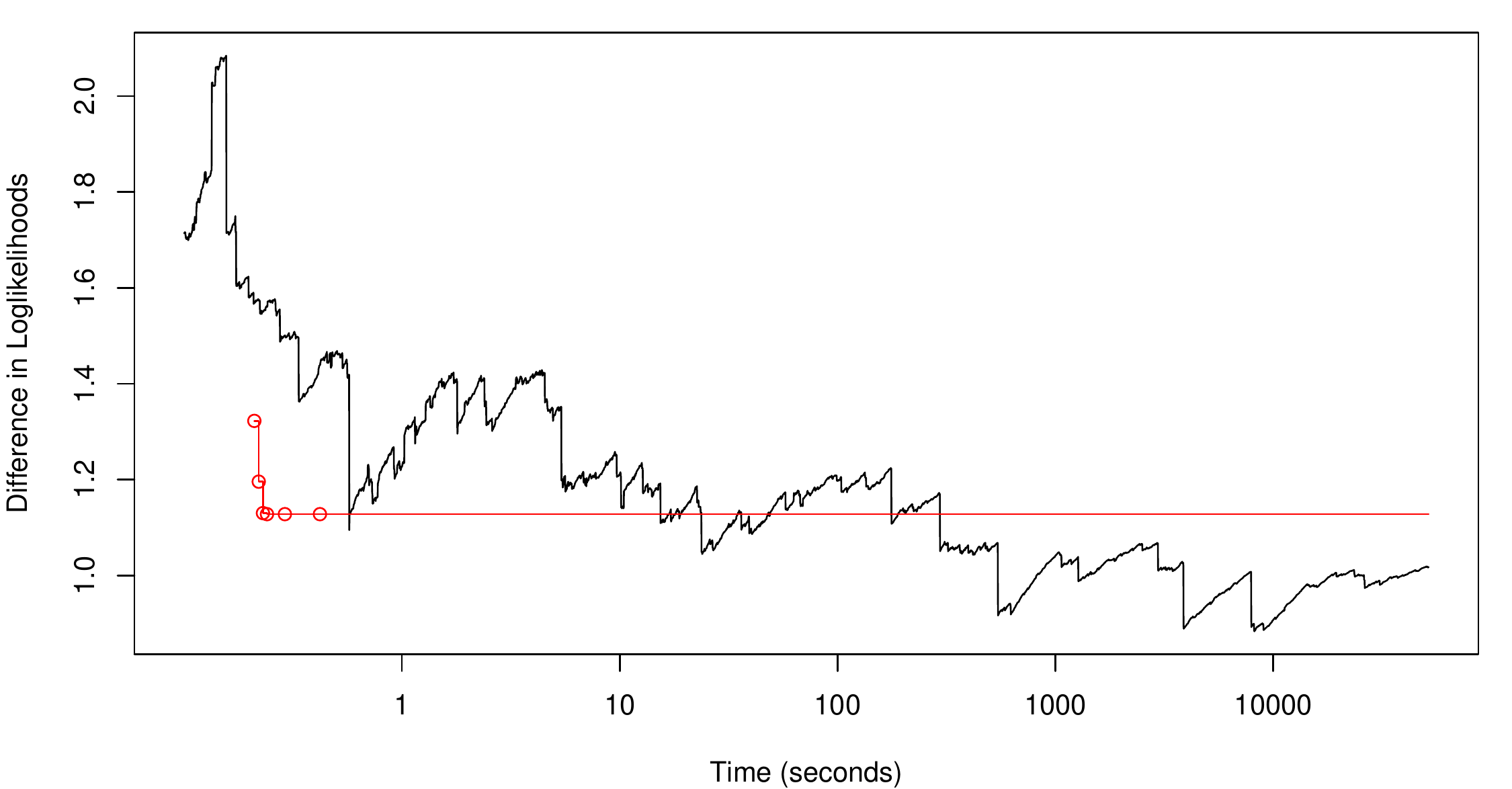}
\caption{\label{fig:l_diff_tree} Importance sampling and sequential reduction approximations to $\ell(0.46,1.30)-\ell(0.60,2.00)$,
plotted against the time taken to find the approximation, on a log scale. The sequential reduction approximation
converges in less than a second, but the importance sampling approximation has still not converged after over 14 hours.}
\end{figure}

We also want to consider the quality of an importance sampling approximation to the
log-likelihood, as described in Section \ref{sec:laplace}. We are interested in the shape
of the log-likelihood surface, rather than the pointwise quality of the approximation, so
we consider approximations to the difference between the log-likelihood at two points:
the maximum $(0.46,1.30)$, and the point $(0.60,2.00)$. We consider the quality of each approximation
relative to the time taken to compute it.
 Figure \ref{fig:l_diff_tree} shows the trace plots of importance sampling and sequential reduction approximations
 to this difference in log-likelihoods, plotted against the length of time taken to find each approximation,
 on a log scale.
  In well under a second, the sequential reduction approximation converges to such an extent that differences
  in the approximations are not visible on this scale. By contrast, after more than 14 hours, the importance sampling approximation
has still not converged.

\subsection{An animal behavior ``tournament'': Augrabies Flat lizards}
\label{example:lizards}

\cite{Whiting_etal_2006} conducted an experiment to determine the factors affecting
 the fighting ability of male Augrabies flat lizards, \emph{Platysaurus broadleyi}.
 They captured $n=77$
lizards, recorded various measurements on each, and then released them and recorded the outcomes of fights between pairs of animals.
The tournament structure is shown in Figure \ref{fig:lizards}.
The data are available in R as part of the BradleyTerry2 package \citep{Firth_Turner_2012}.

There are several covariates $\mathbf{x}_i$ available for each lizard.
\cite{Firth_Turner_2012} suggest to model the ability of each lizard as
$\lambda_i = \beta^T \mathbf{x}_i + \sigma u_i,$
where $u_i \sim N(0,1)$.
The data are binary, and we assume a Thurstone-Mosteller model, so that
$\text{Pr}(\text{$i$ beats $j$}|\lambda_i, \lambda_j) = \Phi (\lambda_i - \lambda_j).$

In order to find the sequential reduction approximation to the likelihood,
 we must first find an ordering
in which to remove the players, an ordering which will minimize the cost of the algorithm.
 Methods
to find upper and lower bounds for the treewidth give that the treewidth is
either $4$ or $5$, and we use an ordering corresponding to the upper bound.

\begin{figure}
\centering
\includegraphics[width=\textwidth]{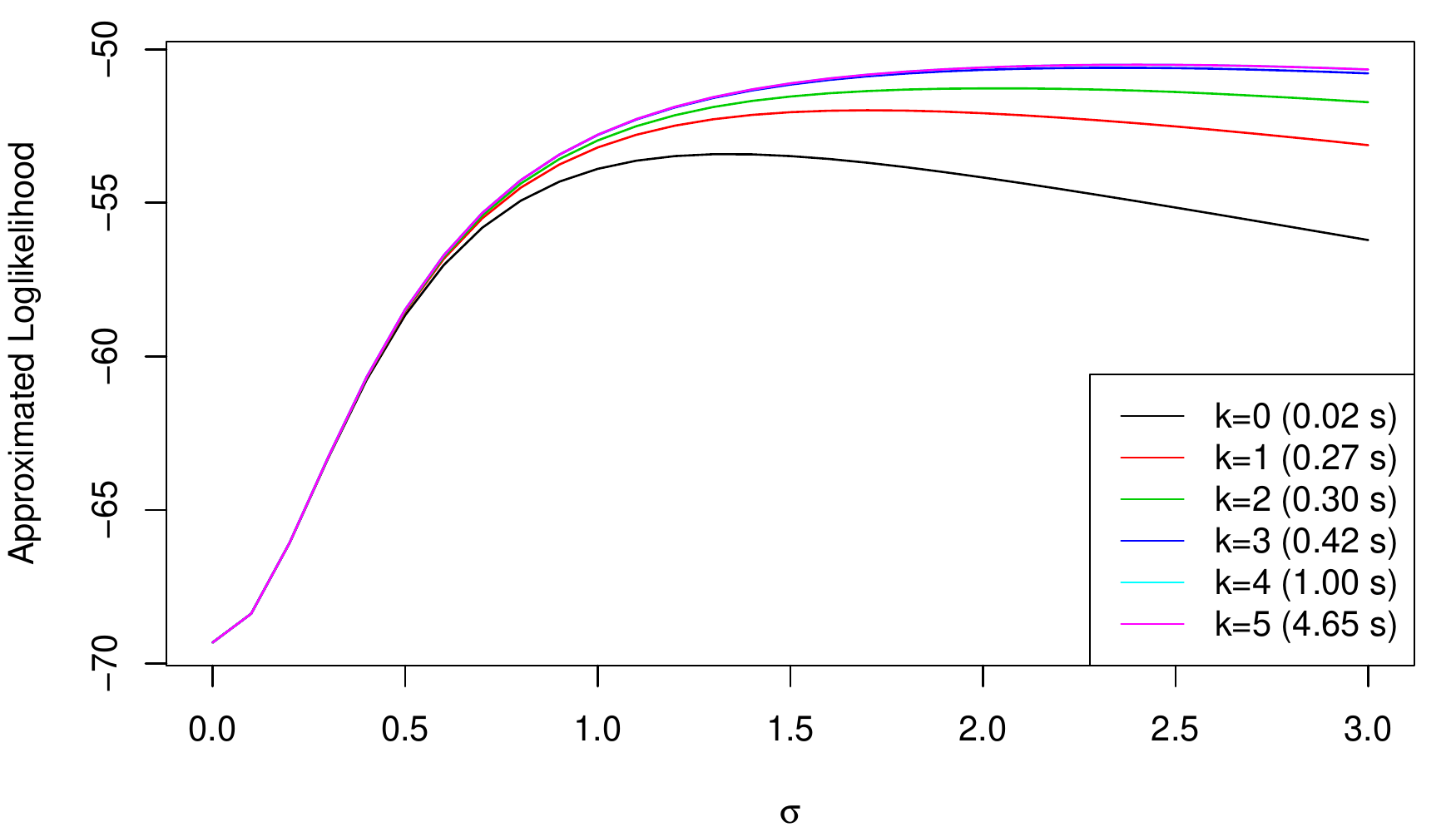}
\caption{\label{fig:lizard_loglikelihoods} Sequential reduction approximations to $\ell(\beta=0,\sigma)$,
for various values of $k$. The curve for $k=0$ (the Laplace approximation) is the lowest line,
and the lines get higher as $k$ increases. The curves for $k=4$ and $k=5$ are indistinguishable.}
\end{figure}

To demonstrate the performance
of the sequential reduction approximation, we consider the cut across the log-likelihood
surface at $\beta=0$, as $\sigma$ varies. The various approximations to this curve are shown in
Figure \ref{fig:lizard_loglikelihoods}.
It becomes harder to obtain a good approximation to the log-likelihood as $\sigma$ increases.
The case $k=0$ corresponds to the Laplace approximation, and gives a poor-quality approximation for $\sigma>0.5$.
As $k$ increases, the approximation improves. All values of $k \geq 3$ give an excellent approximation to the
log-likelihood, and the approximations for $k=4$ and $k=5$ are indistinguishable at this scale.

If we include all covariates suggested by \cite{Firth_Turner_2012} in the model, the maximum likelihood estimator is
not finite. A penalized version of the likelihood
could be used to obtain a finite estimate.
 In a generalized linear model, the bias-reduction penalty of \cite{Firth_1993}
may be used for this purpose.
Further work is required to obtain a good penalty for use with generalized linear mixed models.

\subsection{A three-level model}
\cite{Rabe-Hesketh_etal_2005} note that it is possible to
simplify computation of the likelihood
in models with nested random-effect structure.
Using the sequential reduction method, there is no
need to treat nested models as a special case. Their structure is automatically detected and exploited
by the algorithm.

We demonstrate the method for a three-level model.
Observations are made on items, where each item
is contained within
a level-1 group, and each level-1 group is itself is contained in a level-2 group.
The linear predictor is modeled as
$\eta_i = \alpha + \beta x_i + \sigma_1 u_{g_1(i)} + \sigma_2 v_{g_2(i)},$
where $g_1(i)$ and $g_2(i)$ denote the first and second-level groups to which $i$ belongs.
We consider the case in which there are $100$ second-level groups, each containing two first-level groups,
which themselves each contain two items. The posterior dependence graph of this model is shown in
Figure \ref{fig:three_level}, and has treewidth $2$.
The treewidth of the posterior dependence graph for a similarly defined $L$-level model is $L-1$.

We suppose that $y_i \sim \text{Bernoulli}(p_i)$, where $p_i = \text{logit}^{-1}(\eta_i)$, and
and  simulate from this model, with $\alpha = -0.5$, $\beta=0.5$, $\sigma_1=1$ and $\sigma_2=0.5$,
The fitted values found using the sequential reduction method with various different values
of $k$ are shown in Table \ref{table:three_level_fit}. The parameter
estimates found from the Laplace approximation to the likelihood are some distance
from the maximum likelihood estimator, especially for the variance parameter
of the level-1 group.

\begin{table}
\centering
\caption{\label{table:three_level_fit} The parameter estimates and standard errors for the three-level model}
\begin{tabular}{rrrrrrrr}
  \hline
 & & Laplace & k=1 & k=2 & k=3 & k=4 & k=5 \\
  \hline
\multirow{2}{*}{$\alpha$} & estimate & -0.46 & -0.50 & -0.50 & -0.50 & -0.50 & -0.50 \\
& s.e. & 0.17 & 0.19 & 0.19 & 0.19 & 0.19 & 0.19 \\
 \multirow{2}{*}{$\beta$} &  estimate & 0.45 & 0.49 & 0.49 & 0.49 & 0.49 & 0.49 \\
 & s.e. & 0.23 & 0.25 & 0.25 & 0.25 & 0.25 & 0.25 \\
 \multirow{2}{*}{$\sigma_1$} & estimate & 0.54 & 0.92 & 0.90 & 0.89 & 0.89 & 0.89 \\
 & s.e. & 0.38 & 0.33 & 0.35 & 0.35 & 0.35 & 0.35 \\
 \multirow{2}{*}{$\sigma_2$} & estimate & 0.54 & 0.53 & 0.57 & 0.58 & 0.58 & 0.58 \\
 & s.e. & 0.27 & 0.32 & 0.32 & 0.32 & 0.32 & 0.32 \\
   \hline
\end{tabular}
\end{table}

\section{Conclusions}

Many common approaches to inference in generalized linear mixed models rely on
approximations to the likelihood which may be of poor quality if there is little
information available on each random effect. There are many situations in which it is unclear how good an approximation to the likelihood
will be, and how much impact the error in the approximation will have on the statistical properties
of the resulting estimator. It is therefore very useful to be able to obtain an accurate
approximation to the likelihood at reasonable cost.

The sequential reduction method outlined in this paper allows a good approximation to the likelihood to
be found in many models with sparse structure --- precisely the situation where
 currently-used approximation methods perform worst.
By using sparse grid interpolation methods to store modifications to the normal
approximation used to construct the Laplace approximation,
it is possible to get an accurate approximation to the likelihood for a wide range of models.

\section*{Acknowledgements}
I am grateful to David Firth for helpful discussions. This work was supported by the
Engineering and Physical Sciences Research Council [grant numbers EP/P50578X/1, EP/K014463/1].

\bibliography{sequential_reduction}{}
\bibliographystyle{chicago}
\end{document}